ATP requirements for growth reveal the

bioenergetic impact of mitochondrial symbiosis

William F. Martin

Institute of Molecular Evolution

Heinrich Heine University Düsseldorf

40225 Düsseldorf, Germany

++49-211-81-13011

bill@hhu.de

**Classification**

Biological Sciences, Microbiology

**Abstract**

Studies by microbiologists in the 1970s provided robust estimates for the energy supply and demand of a prokaryotic cell. The amount of ATP needed to support growth was calculated from the chemical composition of the cell and known enzymatic pathways that synthesize its constituents from known substrates in culture. Starting in 2015, geneticists and evolutionary biologists began investigating the bioenergetic role of mitochondria at eukaryote origin and energy in metazoan evolution using their own, widely trusted—but hitherto unvetted—model for the costs of growth in terms of ATP per cell. The more recent model contains, however, a severe and previously unrecognized error that systematically overestimates the ATP cost of amino acid synthesis up to 200-fold. The error applies to all organisms studied by such models and leads to conspicuously false inferences, for example that the synthesis of an average amino acid in humans requires 30 ATP, which no biochemistry textbook will confirm. Their ATP 'cost' calculations would require that *E. coli* obtains ~100 ATP per glucose and that mammals obtain ~240 ATP per glucose, untenable propositions that invalidate and void all evolutionary inferences so based. By contrast, established methods for estimating the ATP cost of microbial growth show that the first mitochondrial endosymbionts could have easily doubled the host's available ATP pool, provided (i) that genes for growth on environmental amino acids were transferred from the mitochondrial symbiont to the archaeal host, and (ii) that the host for mitochondrial origin was an autotroph using the acetyl-CoA pathway.

**Significance statement**

Life is a chemical reaction. It requires energy release in order to proceed. The currency of energy in cells is adenosine triphosphate ATP. Five decades ago, microbiologists were able to measure and understand the amount of ATP that cells require to grow. New studies by evolutionary biologists have appeared in the meantime that brush aside the older microbiological findings, using their own methods to calculate the ATP cost of growth instead. Science is, however, an imperfect undertaking. The new studies contain a major error, similar to conflating centimeters with yards. The error affects many publications and their conclusions. Using the old methods, we can still meaningfully study the role of energy in evolution, including the origin of complex, nucleus-bearing cells.

**Introduction**

Life is an energy-releasing chemical reaction, and energy is the motor of all evolution. Energy in evolution has become the focus of many recent papers concerning the origin of eukaryotic (nucleus-bearing) cells. At the heart of the issue is the question of whether, in mechanistic terms, endosymbiosis or gradualism better account for the origin of eukaryotes and what role mitochondria played therein. Endosymbiosis entails the origin of novel clades via the union of two simpler cells into one, more complex, cell that harbors a new intracellular organelle (mitochondria or chloroplasts). Symbiotic theories imply a stepwise (or quantum) increase in cellular complexity during the prokaryote to eukaryote transition (1, 2) and, in newer formulations, posit an essential role for mitochondrial energy harnessing in bridging the prokaryote-eukaryote divide (3–6). Symbiotic theories for eukaryote origin tend to be mechanistically explicit and are mutually consistent in that most, if not all, of the cellular novelties at the origin of eukaryotes can be recognized as a response to evolutionary pressures caused by the presence of a permanent bacterial endosymbiont in an archaeal host (7). These novelties include spliceosomes (8), the nuclear membrane (9), the origin of the eukaryotic endomembrane system from mitochondrial derived vesicles (9), the Golgi apparatus (7), autophagosomes (10), as well as meiosis and sex (7, 11, 12).

In gradualist theories, mitochondria play no role in eukaryote origin, having no impact—energetic, mechanistic or otherwise—on the emergence of eukaryote complexity (13–23). Gradualist theories operate with classical evolutionary mechanisms including point mutation, gene duplication, ploidy, population size effects, drift and selection rather than symbiotic mechanisms, to generate novel cytological structures and processes that characterize the eukaryotic lineage. They share the common premise that mitochondria played no role eukaryotic emergence, with mitochondria either being absent in the eukaryotic common ancestor altogether (13, 14) or mitochondrial presence in the eukaryote ancestor being a coincidence at best, without causal or energetic effects (15–18, 21, 23). Based on current evidence, either of these mutually exclusive sets of theories could, in principle, be true.

One paper highlighting the bioenergetic role of mitochondria at eukaryote origin (4) figures prominently in this debate. By providing comparative evidence for the bioenergetic significance of mitochondria in eukaryogenesis, a paper by Lane and Martin (4) elicited staunch rebuttal from population geneticists in the form of mathematically detailed and seemingly robust computational constructs by Lynch and Marinov (15), in which the calculated bioenergetic cost of a gene was estimated and presented as hard evidence that mitochondria had no impact on eukaryote origin (15). A series of papers that built upon those calculations (15) followed that unanimously reinforced claims of mitochondrial irrelevance to eukaryote origin (16–18, 21—28). Newer work extends the same variety of bioenergetic calculations to explaining aspects of metazoan evolution (29). These energetically based challenges have, however, brushed aside established knowledge about the ATP requirements for cellular growth in well-studied microbial systems (30–32).

Were the gradualist energetic challenge correct, it would indeed weaken the case for symbiotic theories, begging the question: is it correct? The bioenergetic challenge rests *in toto* upon the original calculations of Lynch and Marinov (15), which have been believed and trusted, but not inspected. The recent claim (29) that 30 ATP are required to synthesize one average amino acid in humans calls stridently, however, for critical inspection of such calculations (15), because it cannot be true and it is not a typo, uncovering instead a recurrent, systematic error that defies textbook biochemistry across a decade of publications, raising two important questions: How large is the error, and does it impact evolutionary inferences contained in the affected papers? Here I report the exact source of error in the calculations of Lynch and Marinov (15), its order of magnitude and its biological implications. Furthermore, I show that using realistically estimated values for ATP growth requirements we can investigate whether the energetics of amino acid and protein synthesis work against or in favor of endosymbiotic theories for eukaryote origin that entail an energetic role for mitochondria (4) involving a methanogenic host (3).

**The cost of synthesizing proteins**

The procedure of Lynch and Marinov (15) calculates the costs, in terms of ATP expense (hydrolysis of high energy phosphate bonds (32) for various cellular processes, as outlined in their 22-page supplement. We focus on only one bioenergetic cost of interest: the cost of synthesizing protein. The reasons to focus on protein are simple and threefold. (i) The main biosynthetic cost that a growing cell encounters is protein synthesis, with peptide bond formation on ribosomes alone comprising about 60% of the energy budget (30). (ii) The cost of synthesizing protein was central to inferences of Lane and Martin (4) regarding the role of mitochondria in fostering eukaryote complexity, which was the main challenge in the report by Lynch and Marinov (15). (3) The cost of synthesizing protein—amino acids specifically—is where a crucial error was incurred that causes their entire computational model (15), and subsequent papers built upon it, to fail.

We start with the composition of the cell, for which *E. coli* is traditionally the standard system of choice. Lynch and Marinov (15) do not specify the protein content for *E. coli* or other cells they model, but they assume the dry weight of an E. coli cell as 0.28 pg/cell, a standard value (~70% water fresh weight). Different studies come to slightly different values for the chemical composition of *E. coli*. Following early reports by Morowitz (33) and Stouthamer (30), the value of 50-55% protein by dry weight can be taken for *E. coli* (**Table 1**). The cost assumed (15) for protein synthesis at the ribosome is uncontested, 4 ATP per peptide bond (30).

**Table 1.** Chemical composition of *Escherichia coli* cells [% by dry weight]

| Cell constituent | Data source (reference) | | |
| --- | --- | --- | --- |
| | Stouthamer (30) | Lengeler (37) | Neidhardt (38) |
| Protein | 52.4 | 50–60 | 55.0 |
| RNA | 15.7 | 10–20 | 20.5 |
| DNA | 3.2 | 3 | 3.1 |
| Lipid | 9.4 | 10 | 9.1 |
| Polysaccharide | 16.6 | | |
| Glycogen | | 2.5–25 | 2.5 |
| Lipopolysaccharide | | 3–4 | 3.4 |
| Murein | | 3–10 | 2.5 |
| Metabolites, ions | | 4 | 3.9 |

Data summarized from references indicated and retabulated from (39).

The costs that Lynch and Marinov (15) use for the synthesis of amino acids are the issue. They calculate that *E. coli* (and all other organisms in their study) expends 23.5 ATP per amino acid for the synthesis of the amino acids from central glucose-derived intermediates such as pyruvate, phosphoenolpyruvate (PEP), 3-phosphoglyerate, erythrose-4-phosphate (E4P), as given in their Supplemental Table 3 (15). By contrast, Stouthamer (30) reports, on average, 1.2 ATP expense for synthesis of an average amino acid (**Table 2** but see also **Figure 1**). How do Lynch and Marinov (15) arrive at a value of 23.5 ATP per amino acid? They use the method of Craig and Weber (34) to calculate amino acid synthesis costs. The Craig and Weber method (34), CW, calculates the cost of synthesizing an amino acid as (i) the number of ATP needed to synthesize the amino acid from universal metabolic precursors *plus* (ii) the amount of ATP that *E. coli could have gained* if it had respired those precursors in $O_2$ instead of making the amino acid *plus* (iii) the amount of ATP that *E. coli could have gained* if it had not invested NADH + $H^+$ or $FADH_2$ into amino acid synthesis, but respired those reducing equivalents in the respiratory chain as well. Craig and Weber (34) assume aerobic growth for these ATP costs and presumably ca. 30 ATP per glucose, though *E. coli* does not pump at complex I in the presence of $O_2$ (35). In essence, the CW method delivers the ATP gain from respiring amino acid synthesis components in human mitochondria, which is why the rightmost column in **Table 2** is included.

The CW method does not deliver ATP 'costs' (**Table 2**), it delivers savings at best because—this is crucial—*E. coli* unconditionally requires amino acids in order to grow. All calculations in Lynch and Marinov (15) and subsequent papers based upon them, assume growth, usually maximum growth rate. If the *E.coli* (or any other) cell is to grow, it needs to double its mass of protein at every cell division, and this condition non-negotiably requires a supply of new amino acids for the new cell equal in mass to the amino acids present in the original cell. By *not* synthesizing the amino acid from glucose and $NH_4^+$ (the savings of the CW method), *E. coli* 'saves' ATP, but it cannot grow. The nitrogen-lacking carbon precursors like oxaloacetate or erythrose-4-phosphate cannot substitute for amino acids at the ribosome to make new cells. There is no conceivable scenario in which *not* synthesizing a required amino acid (the CW method, regardless how calculated) increases or decreases the cost of synthesizing the required amino acid from glucose and $NH_4^+$, or substitutes for the required amino acid. Fully in their defense, Craig and Weber (34) were calculating costs of protein synthesis for colicin plasmids in *E. coli*, not whole cell growth. Clearly, the CW savings method was not designed for application to whole cells, and does not scale accordingly.

The fact that many authors have used the CW method to calculate cell growth energetics in a way that dismisses the salient microbial findings (30–32) on the topic neither remedies the problem nor renders the CW method applicable to estimate amino acid synthesis costs from glucose or other carbon substrate and $NH_4^+$ for growth. Note that the CW method implies that amino acid synthesis would be, on average, 4 times less expensive using the same pathways under anaerobic conditions than under aerobic conditions (**Table 2**), as Wagner (36), who used the CW method, calculated. This factor of 4 is a computational artefact, because under aerobic conditions, less 'savings'—termed 'costs' (36)—are calculated, but no amino acid can be synthesized since the corresponding carbon precursors are either stoichiometrically fermented or respired.

This must be stated clearly, because the error in the calculations of Lynch and Marinov (15) has escaped peer review numerous times: A cell that consumes the precursors for amino acid synthesis via $O_2$ respiration or fermentation can under no circumstances synthesize amino acids from those respired (or fermented) precursor molecules, regardless of ATP supply, because the precursors are converted to excreted waste products, such as $CO_2$, acetate or propionate. In layperson's terms: One cannot build a house with wood that has been burned for heat. The CW method (34) at the foundation of the calculations by Lynch and Marinov (15) cannot be applied to cell growth in any organism.

**Table 2**. Biosynthetic costs, savings, and oxidative ATP yield for amino acids.

| | E. coli grown on glucose, ammonia and salts | | | | | | |
|---|---|---|---|---|---|---|---|
| | Stouth.[a] | Craig and Weber[b] | | Wagner[c] | | Neidhardt[d] | Bender[e] |
| Amino Acid | ATP cost from glucose | ATP cost from glucose | ATP aerobic savings | ATP aerobic savings | ATP anaerobic savings | ATP cost in rich medium[f] | aerobic ATP gain (mitoch.) |
| Ala | −1 | 0 | 12.5 | 14.5 | 2 | 1 | 12.5 |
| Arg | 3 | 7 | 18.5 | 20.5 | 13 | 1 | 25 |
| Asn | 0 | 3 | 4 | 18.5 | 6 | 1 | 12.5 |
| Asp | 2 | 0 | 1 | 15.5 | 3 | 1 | 12.5 |
| Cys | 3 | 4 | 24.5 | 26.5 | 13 | 1 | 12.5 |
| Glu | −1 | 0 | 8.5 | 9.5 | 2 | 1 | 22.5 |
| Gln | 0 | 1 | 9.5 | 10.5 | 3 | 1 | 22.5 |
| Gly | 0 | 0 | 14.5 | 14.5 | 1 | 1 | 12.5 |
| His | 7 | 6 | 33 | 29 | 5 | 1 | 22.5 |
| Ile | 1 | 2 | 20 | 38 | 14 | 1 | 34 |
| Leu | −3 | 0 | 33 | 37 | 4 | 1 | 33 |
| Lys | 0 | 2 | 18.5 | 36 | 12 | 1 | 22.5 |
| Met | 4 | 7 | 18.5 | 36.5 | 24 | 1 | 31.5 |
| Phe | 2 | 1 | 63 | 61 | 10 | 1 | 29 |
| Pro | 0 | 1 | 12.5 | 14.5 | 7 | 1 | 27.5 |
| Ser | 0 | −1 | 15 | 14.5 | 1 | 1 | 12.5 |
| Thr | 2 | 2 | 6 | 21.5 | 9 | 1 | 19 |
| Trp | 5 | 5 | 78.5 | 75.5 | 14 | 1 | 37.5 |
| Tyr | 2 | 1 | 56.5 | 59 | 8 | 1 | 31.5 |
| Val | −2 | 0 | 25 | 29 | 4 | 1 | 27.5 |
| Sum | 24 | 41 | 472.5 | 581.5 | 155 | 1 | 431.5 |
| Avg. | 1.2[g] | 2 | 23.6 | 29.1 | 7.7 | 1 | 21.5 |

Sources: [a] ref. (30), [b] ref. (34), [c] ref. (36), [d] ref. (38). [e] ref (39). The rightmost column shows the amount of ATP that humans can obtain from respiring amino acids. [f] Stouthamer (30) calculates approx. one ATP per amino acid for import across the plasma membrane (or ammonia import in the case of minimal media). ATP is generated from glucose in the process of generating some carbon precursors in *E. coli*, hence some amino acids have a negative cost (net ATP gain) in synthesis from glucose and ammonia (30). [g] The true cost of an 'average' amino acid in *E. coli* has to be weighted against the frequency of the amino acid in its proteins, see **Figure 1**. Though quantitatively less serious, the same problem discussed in the present paper for amino acids is also

encountered for nucleic acids, because Lynch and Marinov (15) calculate, and use for every organism, a 'cost' of 50 ATP per polymerized base (100 ATP per base pair), whereby the cost of synthesizing an average base incorporated into nucleic acid in *E. coli* is 7.5 ATP from glucose and $NH_4^+$ (30), not 50 ATP.

**How much do amino acids actually cost?**

The source of most microbial ATP cost estimates for growth trace to the paper of Stouthamer (30), who tabulated the chemical composition of the cell, the biosynthetic costs for cell growth from biosynthetic pathways and—crucially—vetted those numbers against laboratory growth yield experiments. **Table 3** summarizes the values tabulated by Stouthamer (30) for *E. coli* aerobic growth on several different media using the classical values of dry weight composition for *E. coli* cells from Morowitz (33).

From **Table 3**, the synthesis of 524 mg protein in 1 gram dry weight of *E. coli* grown on rich medium (containing all amino acids and nucleic acid bases) requires 19.1 mmol ATP out of the total of 31.4 mmol ATP required to synthesize a gram of cells on rich medium. The peptide bond synthesis reaction at the ribosome thus corresponds to 61% of the total ATP expenditure of the cell. The pure cost of synthesizing peptide bonds is 4 ATP each: 2 from $PP_i$ formation at aminoacyl tRNA synthesis, which renders the reaction irreversible (42, 43), and 1 GTP each for the two elongation factors. For growth on rich medium, there is no cost for synthesizing amino acids, but there is a cost for their import (**Table 3**). Does 19.1 mmol ATP for protein per g of cells add up? Yes. At a cost of 19.1 ATP for peptide bond formation and 4 ATP per peptide bond, the cell has 4.78 mmol of peptide bonds or 4.78 mmol of amino acids with an average molecular weight of 110 g/mol each (a standard biochemical conversion) yielding 0.523 g of amino acids per cell, corresponding to 52.4% dry weight protein in 1 g of cells (**Table 3**). That was for rich medium supplied with amino acids and bases.

**Table 3**. ATP requirement of per gram of cells under growth on different substrates.

| | | | ATP requirement (mmol ATP per g dry weight) by medium | | | | |
| | | | minimal medium (inorganic salts) and O$_2$ plus | | | | |
| | g/g | rich | glucose | lactate | malate | acetate | CO$_2$ |
|---|---|---|---|---|---|---|---|
| Synthesis of: | | | | | | | |
|   Protein | 52.4% | 19.1 | 20.5 | 33.9 | 28.5 | 42.7 | 90.7 |
|   RNA | 15.7% | 3.8 | 5.9 | 8.5 | 7.0 | 10.1 | 21.2 |
|   DNA | 3.2% | 0.58 | 1.05 | 1.6 | 1.3 | 1.9 | [b] |
|   Polysaccharide | 16.6% | 2.05 | 2.05 | 7.1 | 5.1 | 9.2 | 19.5 |
|   Lipid | 9.4% | 0.15 | 0.15 | 2.7 | 2.5 | 5.0 | 17.2 |
| Transport | n.a. | 5.75 | 5.21 | 20.0 | 20.0 | 30.6 | 5.2 |
| Total | 97.3[a] | 31.5 | 34.8 | 73.8 | 64.4 | 99.5 | 153.8 |

Data from Stouthamer, 1977 (31). [a] ignores ca. 3% metabolites and salts (30). [b] The value of 21.1 for RNA includes DNA. Using the CW method of amino acid synthesis cost, it has been estimated that *E. coli* requires 20-50 billion ATP to synthesize a new cell (40). However, if one uses the values provided by Stouthamer (30), which square off well with growth yields per mol ATP synthesized for *E. coli* (31) and other cells (32, 41) the ATP requirement to build a *E. coli* cell on rich medium or minimal medium with glucose and ammonia, is roughly (31.5 mmol ATP per g of cells) × (0.3 × 10$^{-12}$ g per cell) × (6.02 × 10$^{23}$ per mol) = ca. 5.7 billion ATP per cell division, plus growth rate dependent allocations for maintenance energy (30, 32). The roughly 3–9-fold elevated estimate of total ATP requirement for synthesizing an *E. coli* cell (40) results from using the CW method to calculate biosynthetic costs. Independent from this study, Ortega-Arzola et al. (50) noted that the calculations of Lynch and Marinov (2015) deliver ATP requirements for synthesizing an *E. coli* cell that exceed estimates based on the free energy of cell formation. They assumed, however, that the calculations of (Lynch and Marinov 2015) were valid, which is not the case.

As explained by Stouthamer (1973), the additional cost of synthesizing amino acids for 1 g of cells is obtained by simply subtracting 19.1 from 20.5 (growth on glucose and ammonium) = 1.4 mmol ATP for amino acid synthesis from glucose. Does that add up? Yes, however, the cost of 1.4 mmol ATP per 4.78 mmol of amino acids averages to only 0.29 ATP per amino acid, less than the unweighted average (1.2 ATP per amino acid) in the first column of **Table 2**. The apparent discrepancy resides in the fact that synthesis of the amino acids most commonly used by *E. coli* have no cost, instead they generate a small net ATP gain from glucose as calculated by Stouthamer (30), who explains the exact source of the ATP gains from amino acid synthesis from glucose on p. 544 of his freely available paper. The biosynthetically most expensive amino acids in *E. coli* are rare (**Figure 1**), the most common ones deliver ATP gains from glucose. If the ATP costs of amino acid synthesis (taking into account ATP gains) are weighted by the frequency of amino acids in *E. coli* dry matter—given in Table 4 of (30)—the synthesis of an average amino acid from glucose costs 1.36 mmol ATP per 4.78 mmol of amino acids, or average 0.28 ATP per amino acid, which explains the discrepancy relative to the unweighted average of 1.2 in **Table 2**.

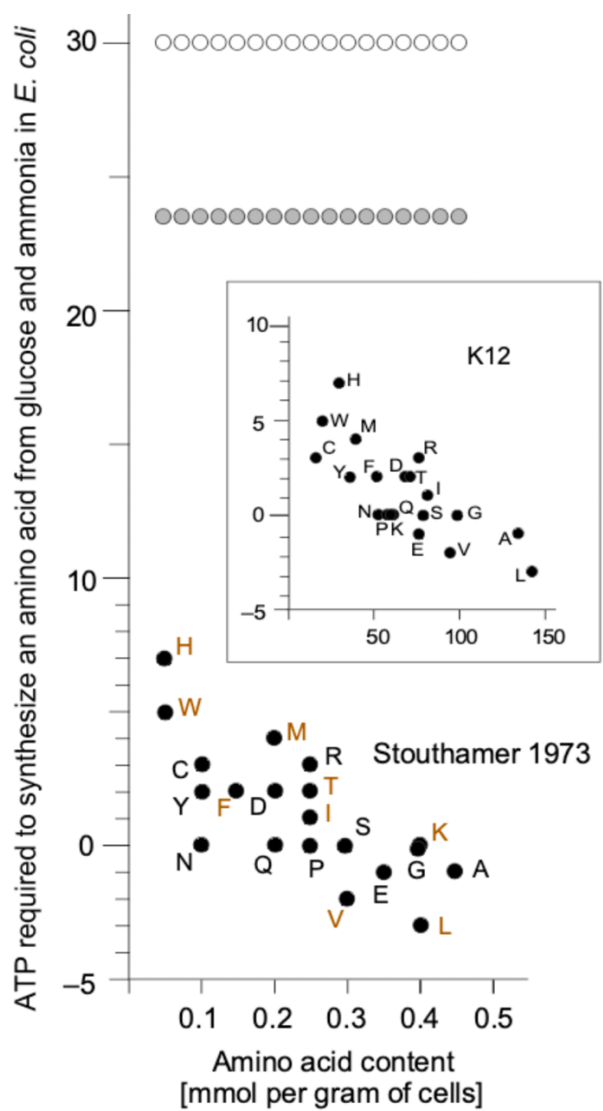

**Figure 1**. Cost of synthesizing an amino acid in *E. coli* versus amino acid frequency. Black circles indicate the values from Stouthamer (30), who used the amino acid content values reported by Morowitz (33). Gray circles at 23.5 ATP indicate the values

for all organisms used by Lynch and Marinov (15), white circles at 30 ATP indicate the values for humans from Lynch (29). Amino acids shown in sepia font indicate the essential amino acids that mammals and most animals cannot synthesize and hence have to be obtained from food; essential amino acids are among both the most and least expensive to synthesize. The inset shows the same amino acid synthesis costs as in the main panel (Y-axis) but plotted against the frequency of amino acids specified in the *E. coli* K12 genome sequence (X-axis, relative units) for comparison.

**Does it make a difference?**

The original report by Lynch and Marinov (15), and subsequent papers that use their method, does not differentiate between aerobic growth or anaerobic growth, amino acid import in food (as in metazoans) or synthesis of all amino acids. Instead they use one cost, 23.5 ATP per amino acid that ends up in protein, regardless of how that amino acid was obtained, for example from food, with $O_2$, without $O_2$, using photosynthesis, the Calvin cycle, the acetyl-CoA pathway, diazotrophy, or other metabolism for organisms in their study. The 23.5 ATP cost is at least 84 times higher (23.5/0.29) than the actual biosynthetic cost for *E. coli*. The value of 23.5 is also 6 times higher than the cost of 4 ATP per peptide bond at the ribosome, which consumes 61% of the cell's energy (see **Table 3**). This is important: At the cost of 23.5 ATP per amino acid (15), *E. coli* would be consuming (non-negotiable) 4 ATP per peptide bond (19.1 mmol ATP per g) plus 23.5 ATP per peptide bond (each amino acid). That is, it would be investing not 19.1 mmol ATP per g but 131 mmol ATP per g of cells, which is 377% of its actual energy requirement per cell division (30, 31). Because the amount of glucose it consumes during growth is a known, measured value (30, 31), *E. coli* would have to be obtaining 96 ATP per glucose through respiration using the Lynch and Marinov (15) model—an absurd proposition. The cost of amino acids makes a difference.

In the most recent paper (29), the value of 23.5 increases to 30 ATP per amino acid incorporated into proteins for organisms that do not synthesize half of their amino acids and in carnivorous mammals that are specialized to a protein diet. Because essential and nonessential amino acids are evenly distributed across frequency for *E. coli* (**Figure 1**), an amino acid biosynthesis in a mammal can cost up to a maximum value of roughly 0.14 ATP per amino acid, conservatively assuming that no non-essential amino acids from protein diet are incorporated into protein synthesized (39), such that for mammals, the biosynthetic cost of amino acids is overestimated (29) by 30/0.14, a factor exceeding 200.

That too, makes a difference. As an example, it is reported (29) that the cost of synthesizing a human mitochondrial ATP synthase consisting of 5380 amino acids is 183,000 ATP per ATP synthase protein complex, calculated as 5380 × 30 = 161,400 ATP for amino acid synthesis plus 21250 ATP for peptide bond formation. The realistically estimated cost of synthesizing the same ATP synthase is 5380 × 4 ATP per peptide bond = 21250 ATP, plus (maximum) 858 ATP (the cost of synthesizing the 11/20 non-essential amino acids in the complex assuming they are not incorporated from food, at roughly 0.29 ATP each (30) (**Figure 1**)) for a total cost of roughly 22,108 ATP per ATP synthase complex. The remaining 160,000 ATP, 88% of the 'cost' calculated per ATP synthase (29), do not exist in nature, they are a computational product of the CW method that was used (29) to calculate amino acid synthesis costs in all organisms. Calculating the cost of synthesizing the 5380 amino acids for an enzyme as 161,400 ATP when the true cost is approximately 858 ATP increases the 'cost' of synthesizing an ATP synthetase (29) over the true cost by 183000/22108 = 8.3, roughly an order of magnitude.

Does it make a difference? Consider the impacts for food webs or ecology and evolution, where these calculations are being applied (29). Animals consist of 60-80% protein dry weight depending on the species and growth conditions (44), whereby agriculturally important land animals typically consist of >80% protein by dry weight (45). According to (29), a cow would have to supply over 183000/22400 = 8 times more ATP per cell (a roughly 8-fold increased food uptake and respiratory rate at a constant ~30 ATP per glucose) than it actually does in order to grow at observed rates—grow means synthesize protein. That means that a cow, at a constant ATP gain per gram of food, would need to eat 8 times more food, by mass, per unit time than the real-world value in order to gain weight at observed rates. Cows can gain weight at a rate of about 1 kg per day (46), about 80% of that weight gain (by dry weight) is protein. Weight gain requires food. Under modern conditions, about 3–10 kg of maize have to be fed per kg of beef formed (47). According to the model of cellular energetics in which animals consume 30 ATP per amino acid synthesized as protein (29), a cow would need to be eating 24–80 kg of maize per day to make one kg of beef. On real farms, about 5 kg are sufficient (47). Alternatively, at known food intake rates and 30 ATP to synthesize an amino acid, the cow's mitochondria would have to be obtaining roughly 240 ATP per glucose, rather than the well-vetted value of ~30 ATP per glucose (48, 49) in order to satisfy the published (29) calculations. In an ecological or evolutionary context, models assuming 8-fold inflated biomass growth energetics would have each trophic level requiring 8 times more food than the previous, because all organisms need to synthesize protein, regardless their size, or it would have animals synthesizing 8 times more ATP per glucose than their mitochondria can deliver (48, 49). The value of 23.5 that is incorrect for *E. coli* is also incorrect for the elephant.

All analyses and correlations of Lynch and Marinov (15) that involve the ATP requirements of cells, the energy budget of cells, or correlations between energy budgets and growth rate are fundamentally in error. All subsequent papers that use their method to draw inferences are equally void.

**Symbioses of cells with identical physiologies yield competition, not benefit.**

The present findings show that growth associated ATP cost calculations (15) used to counter symbiotic models of eukaryote origin (4) fail because the most important bioenergetic cost of the cell, protein synthesis, was overestimated by a factor of 8, whereas the ATP synthesis rate was kept at real values. No cell in any of their models would be able to grow at observed rates with such a budget. Their inferences that trivialize the energetic benefit of mitochondria fail accordingly. This is important because Schavemacher and Munoz-Gomez (26) recently used the same method (15) to investigate the energetics of eukaryote origin by modelling a symbiosis involving a host cell without endosymbionts in comparison a host cell to a host cell with mitochondrial ATP synthesis. Their findings (26), like those of the earlier study (15), uncovered no energetic impact of mitochondria at eukaryote origin, and have been brightly advertised as evidence in favor of gradualist models (27), or against symbiotic models, or both. Their study modeled a wide range of cell sizes and estimated symbiont costs with respiratory deficits and other variables (26). In all cells and all conditions SM-G modeled, the host and the symbiont were, however, seen from the physiological and energetic standpoint, (i) respiring glucose with $O_2$, (ii) respiring $O_2$ at their plasma membrane, (iii) always synthesizing proteins at 23.5 ATP per amino acid (plus 4 ATP per amino acid for translation at the ribosome), and (iv) both cells were heterotrophs.

The reason that such studies (15, 16, 21, 22, 26) find no difference between prokaryotes and a eukaryote with mitochondria is that all cells in their models have exactly the same metabolism, using exactly the same substrates and experiencing exactly the same inaccurately calculated costs. That underscores a faulty premise common to gradualist approaches to eukaryote origin: if the host is heterotrophic (2, 13–19, 21–23) it has no need for an endosymbiont, on the contrary, host and symbiont will compete for the same resources rather than enter into a symbiosis (3, 5, 51). There are of course symbioses known where both partners are heterotrophic, for example the endosymbiotic bacteria of insects (52). But in those highly derived bacterium-animal symbioses, the benefits are nutritional, not energetic, in that each partner reciprocally synthesizes only half of the 20 amino acids, namely those needed by the other partner (53). In a symbiosis involving cells with identical heterotrophic, respiring physiology (15, 26), it is indeed difficult to identify an energetic difference with and without mitochondria. That is why microbial symbioses in the real world typically involve cells with fundamentally different energy metabolisms, such that tangible energetic benefit from physical association and symbiosis accrues (54, 55).

**What if the host was a methanogen?**

The central issue remains: Do energetics favor a role for mitochondria at eukaryote origin (4) or not? We can use the present insights to revisit the energetics of the hydrogen hypothesis (3), which posits that the host was a methanogen, and where the metabolisms of the host and symbiont are very different and based on anaerobic syntrophy (54). This requires estimating the values for the cost of amino acid synthesis in a methanogen, because the values given by Stouthamer (31) for an autotroph (**Table 3**) are for the Calvin cycle (the only $CO_2$ fixing pathway well-known at the time), which is energetically expensive in terms of ATP synthesis, 7 ATP per pyruvate synthesized from $CO_2$ (56). As outlined in **Figure 2**, methanogens use the acetyl-CoA pathway, which starts from $H_2$ and $CO_2$ and generates both acetyl-CoA and pyruvate without ATP investment (57), such that these carbon backbones have a net cost of 0 ATP each, as do C1 intermediates of the acetyl-CoA pathway. The reason for this energetically favored $CO_2$ fixation is that in the reaction of $H_2$ with $CO_2$ under anaerobic conditions, the equilibrium lies on the side of reduced carbon compounds (32, 58, 59). Succinyl-CoA could, in principle, also be counted at a cost of 0 ATP because of ubiquitous

acetate:succinate CoA transferases (60), but the reaction in autotrophic metabolism consumes one ATP (or GTP) per succinyl-CoA synthesized (61), probably for thermodynamic reasons, and is counted accordingly. Methanogen ATP synthesis generates 0.5 ATP per methane and does not require concomitant carbon or nitrogen assimilation (41,72). The cost of synthesizing the key intermediates for amino acid biosynthesis from $H_2$ and $CO_2$ in a hydrogenotrophic methanogen (**Figure 2**) are 2 ATP for phospho*enol*pyruvate, 2 ATP for 3-phosphoglycerate, 3 ATP for oxalacetate, 4 ATP for 2-oxoglutarate, 3 ATP for sugar phosphates, 5 ATP for PRPP. Those are the costs of the carbon backbones, but amino acids contain nitrogen.

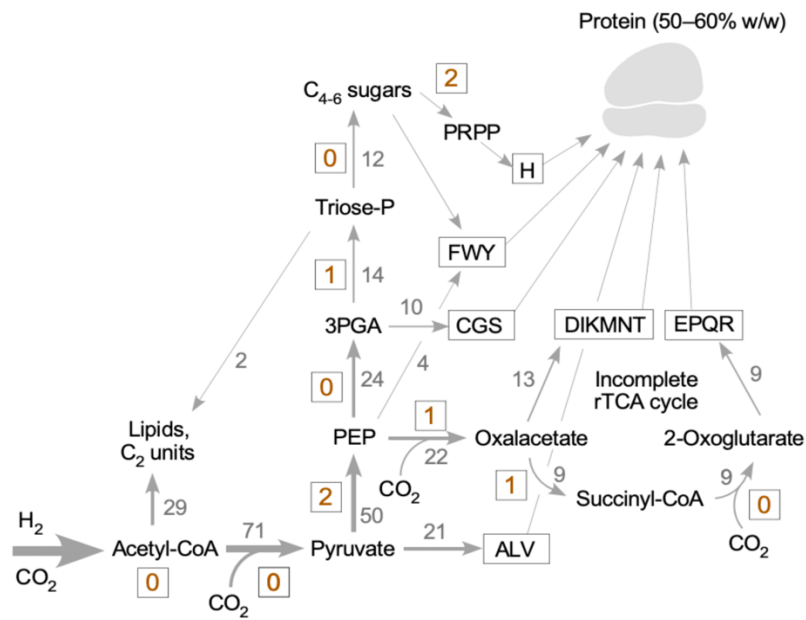

**Figure 2**. Cost of synthesizing carbon backbones for amino acid synthesis in an idealized hydrogenotrophic methanogen. The ATP expense is given in boxed sepia numbers next to conversions. An arrow can indicate several enzymatic steps, width of arrows symbolizes flux amounts. Gray numbers next to arrows indicate the relative flux of carbon from 100 C2 units on acetyl-CoA to key intermediates in organisms that use the acetyl-CoA pathway for $CO_2$ fixation as given by Fuchs (57). The amino acid biosynthetic families are given in one letter code (62) and boxed. The figure is modified from (59).

We can consider two possibilities concerning nitrogen metabolism: the host used $NH_4^+$ like *E. coli*, or was $N_2$-fixing (diazotrophic) like *Methanococcus thermoautotrophicus* (63,64). For $NH_4^+$, one ATP is required for each $NH_4^+$ incorporation into amino acids at the glutamine synthase reaction (1 ATP), the remaining reactions distributing N across metabolism involve reductive aminations (65) or transaminations (66), which consume no ATP (37,67). For $N_2$-fixation, which synthesizes 2 $NH_4^+$ at the expense of 16 ATP (68), an additional cost of 8 ATP per nitrogen atom in organic compounds is incurred. The cost calculations for the synthesis of amino acids in *Methanococcus* are given in supplemental **Table S1**. The cost estimates for synthesizing one gram of *E. coli* or *Methanococcus* cells are shown in **Table 4**.

**Table 4**. ATP costs of *E. coli* vs. *Methanococcus* by nitrogen source

|  |  | \multicolumn{2}{c}{ATP requirement (mmol ATP per g)} |  |  |
| --- | --- | --- | --- | --- | --- |
|  |  | *E. coli* | | *Methanococcus* | |
|  | g/g | rich | glc-$NH_4^+$ | $NH_4^+$ | $N_2$ |
| Synthesis of: | | | | | |
|   Protein | 52.4% | 19.1 | 20.5 | 43.4 | 93.5 |
|   RNA | 15.7% | 3.8 | 5.9 | 13.0[a] | 34.6[a] |
|   DNA | 3.2% | 0.58 | 1.06 | 2.0 | 5.2 |
|   Polysaccharide | 16.6% | 2.05 | 2.05 | 5.1 | 5.1[b] |
|   Lipid | 9.4% | 0.14 | 0.14 | 4.0 | 4.0[c] |
| Total (synthesis)[e] | 97.3 | 25.7 | 29.7 | 67.5 | 142.4 |

Notes: Values for *E. coli* are from Stouthamer (31). The costs of 20 amino acid syntheses for *Methanococcus thermolithotrophicus* are calculated in supplemental **Table S1.** Any $H_2$-dependent hydrogenotrophic methanogen capable of diazotrophic growth could be used in this example, hence use of the general term *Methanococcus* here. For convenience we assume the same g/g chemical composition for the bacterium and the archaeon. [a] For nucleotide synthesis on $NH_4^+$ in *Methanococcus* the costs of precursors (in ATP) given in the text are taken from Lengeler *et al.* (1999). For growth on $N_2$, add 8 ATP per nitrogen atom in the final monomer. The *Methanococcus thermolithotrophicus* genome is 1.7 Mb, smaller than *E. coli*, but its copy number is not specified here, we assume cell size and DNA content of *E. coli*.
[b] Polysaccharide synthesis from glucose in *E. coli* costs 2 ATP per glucose polymerized (30). Glycogen synthesis in *Methanococcus* costs 3 ATP per glucose-P plus 2 ATP for polymerization as UDP-glucose synthesis is $PP_i$-forming (69) or 5 ATP per glucose polymerized. We assume 16.6% dry weight polysaccharides for *Methanococcus*, which is approximate but not unrealistic, as glycogen is present in *Methanococcus*

*thermolithotrophicus* as 13% of protein content or about 7% dry weight (70) and can be present in the same amount as protein in some archaea (71); the methanogen S-layer consists of glycoprotein. Assuming 16.6% polysaccharides has the convenience that multiplying the *E. coli* ATP requirement by the ratio of costs in *E. coli* and the archaeon (5/2) obtains the archaeal value. <sup>c</sup> *Methanococcus* uses the mevanolate pathway to form C5 units from acetyl-CoA, which requires 3 ATP per C5 unit or 12 ATP per phytanyl unit. Synthesis of glycerol-P from $H_2$ and $CO_2$ requires 3 ATP, or 27 ATP per phospholipid monomer. We assume for convenience 9.4% dry weight lipids for *Methanococcus*. <sup>e</sup> Stouthamer (30) calculates roughly 5 mmol ATP per g of cells for transport in addition, mostly for import of amino acids or $NH_4^+$. $N_2$ diffuses across membranes without transport.

---

The hydrogen hypothesis posits that the eukaryotes arose from anaerobic syntrophy between a facultatively anaerobic bacterium (the symbiont) and a $H_2$-dependent autotrophic archaeon (the host). Anaerobic syntrophy is widespread in nature and is generally understood in terms of bioenergetics (54, 55): $H_2$ and $CO_2$ produced from ATP synthesis via substrate level phosphorylation during bacterial fermentations are growth substrates for $H_2$-dependent methanogens, which obtain their carbon via the acetyl-CoA pathway (**Figure 2**) and their ATP from methanogenesis, generating 0.5 mol of ATP per methane (72). Methanogens cannot grow from glucose (73) or carbon substrates larger than pyruvate (74). The $NH_4^+$ required for amino acid synthesis is either imported as $NH_4^+$ or they are diazotrophic, fixing $N_2$ in the cytosol via nitrogenase (76). Gene transfers from the mitochondrial endosymbiont to the archaeal chromosomes of the host (3, 4, 9) imprint the metabolism of the endosymbiont onto the chromosomes and cytosol of the host, transforming an $H_2$-dependent, autotrophic host into a heterotroph harboring a facultatively anaerobic organelle, the common ancestor of mitochondria and hydrogenosomes (77).

Amino acid metabolism has energetic impact on that symbiosis. Because cells are 50% protein, proteins are the most common substrates for fermenters in deep sea marine environments (78), the environment where the hydrogen hypothesis was set. Amino acid fermentations typically involve deamination to the corresponding 2-oxoacid, which undergoes decarboxylation to form an acyl-CoA thioester that is converted to an acyl phosphate for ATP synthesis (38). The end products of the fermentation are an organic acid, $H_2$, $CO_2$, and $NH_4^+$, with $H_2$, $CO_2$ and $NH_4^+$ (and possibly acetate) being initial growth substrates for the host. If the symbiont transferred genes for amino acid importers to the host, and if they became expressed in its plasma membrane, the symbiont would thereby enable the host compartment to import amino acids from the environment for protein synthesis rather than having to synthesize them itself. This simple rearrangement of preexisting components (genes and proteins) via endosymbiotic gene transfer (3) has a substantial bioenergetic impact: The host compartment still has to expend 19.1 mmol ATP per g of cells for peptide synthesis, but 24.3 mmol ATP per g (43.4–19.1) are no longer required for amino acid synthesis, ATP that is liberated for other reactions. The amount of ATP liberated (24 mmol ATP per g) is approximately that required to synthesize a cell's worth of protein (19 mmol per g) at the ribosome. But with the bipartite cell's energetic problems solved, thanks to mitochondria (3, 4), the host compartment is not constrained to synthesize more bioenergetic machinery or ribosomes, it has ATP available in amounts that would allow it to synthesize novel, bioenergetically immaterial proteins and thus explore protein sequence space.

This is the crux of Lane and Martin's (4) energetic proposal: Mitochondria do not simply supply more ATP to make cells become bigger (14, 26), they enable the cell to do more of its most expensive and creative evolutionary task: express protein, hence invent novel proteins and functions specific to the (complexity of) the eukaryotic lineage (4). Such evolutionary invention is vetted and filtered by selection and thus comes at a trial-and-error energetic cost, which gradualist theories miss (15-29). In order to explore protein sequence space, the cell requires ATP in amounts that allow exploratory protein synthesis at no penalty (6). That is, the host compartment can experiment with overexpressing structural proteins such as prokaryotic actins or tubulins, the latter for chromosome division (11), in addition to expressing proteins that generate shape and modulate membrane flux (7, 9). That differs from simply making more of the same proteins leading to larger cell size (15, 26). Grown on $NH_4^+$, the energetic benefit of mitochondrial symbiosis (4, 6) incurred from amino acid metabolism, 24 mmol ATP per g, is sufficient to synthesize cell's worth of exploratory proteins while generating the required copy of the original cell's protein content (19 mmol ATP per g) (**Table 4**).

If the host was $N_2$-fixing (**Table 4**), the amount of ATP liberated by importing amino acids as opposed to synthesizing them from $H_2$, $CO_2$ and $N_2$ increases further to 74.4 mmol ATP per gram of cells (93.5 – 19.1), enough to synthesize roughly 4 cell's worth of peptide bonds on (archaeal) cytosolic ribosomes. That is a very substantial amount of liberated, uncommitted ATP that could fuel the exploration of protein structural space and forge protein-based novelties that were present in the eukaryote common ancestor and that are specific to the eukaryotic clade. For those still in search of an energetic benefit for mitochondria (15-29), heterotrophy is yet one more.

The present example of amino acid synthesis underscores energetic advantages of mitochondrial symbiosis that only become manifest if the host is an autotroph and if costs are calculated in accordance with physiology (30-32). If a postulated transition from chemolithoautotrophy to heterotrophy at eukaryote origin was evolutionarily advantageous, did other lineages of methanogens undertake a similar physiological transition? Possibly. Archaeal halophiles are transformed methanogens that acquired a large donation of genes from a bacterial donor to convert them from strictly anaerobic, $H_2$ dependent autotrophs into $O_2$ dependent heterotrophs, yet without formation of a bacterial organelle (79). The origin of archaeal halophiles, which thrive on very salty peptone-rich media, mirrors that of eukaryotes in a physiological and energetic context, yet without the fixation of a mitochondrion equivalent and without the product of the symbiosis having attained eukaryote complexity. Halophiles did not evolve along a trajectory that led to cellular complexity. What did they do with their ATP surplus during their transition to heterotrophy? Archaeal halophiles are conspicuously polyploid, with some species harboring in excess of 20 copies of the genome per cell (80). While DNA synthesis in *E. coli* grown on $NH_4^+$ is not expensive, if multiplied by 20 per cell, the ATP cost of DNA in halophile increases to the level required to make a cell's worth of peptide bonds (**Table 4**). That is an energetic cost that a methanogen-turned heterotroph could readily afford, either for synthesizing new proteins or, alternatively, to bask in the luxury of 20 genomes, when one would suffice. Pronounced polyploidy in archaeal halophiles could be a relic of the energetic advantage conferred by the origin of heterotrophy (79) in their lineage.

**Conclusion**

The issue here is whether mitochondrial energetically contributed to eukaryote origin, or not. The answer is that (i) it depends on whether the ATP costs of growth are calculated in such a way that the energy budget, cell mass and growth add up, which Stouthamer (30) did, Lane and Martin (4) did, but Lynch and Marinov (15) did not, and (ii) it depends on what kind of a symbiosis one models at eukaryote origin. A heterotrophic host has no need for a heterotrophic mitochondrial symbiont (3, 51). There is currently much excitement about archaeal clades inferred from metagenomic data that possess some interesting genes related to eukaryotic cell biological functions, and that are being considered as models for the host of mitochondria at eukaryote origin (81, 82). However, only two such archaea have been cultured so far. They are, like the famous spaghetti-shaped *Korachaeon cryptofilum* isolated by Stetter (83), amino acid fermenters (84, 85), but with an interesting appendage-producing morphology that (i) probably serves to increase surface area for substrate acquisition and that (ii) was previously observed in other archaeal fermenters (86, 87).

Since Margulis's day (2), all models for the origin of eukaryotes assume that the host for the origin of mitochondria was heterotroph, with one exception (3, 55). For a heterotroph, in particular an amino acid fermenting archaeon, there is indeed little energetic benefit to be construed from acquiring a mitochondrion. By contrast, a methanogen that drifts away from a geological source of $H_2$ (3) unconditionally needs its $H_2$-producing symbiont to survive. The new clades of archaea that branch near eukaryotes in phylogenetic trees all seem to be derived from methanogens, in a phylogenetic sense, and it is possible if not likely that all archaea are derived from methanogens to begin with (58, 88–92). It is thus well within the realm of microbial reason that the host cell at eukaryote origin was a methanogen. Methanogens present favorable symbiotic partners for the origin of mitochondria (3), as the latter can substantially improve the energetics of the former (4) through endosymbiosis.

**Acknowledgements**
I thank Sven Gould, Parth Raval and John F. Allen for critical and helpful comments.

**Funding**
This project has received funding from the European Research Council (ERC) under the European Union's Horizon 2020 research and innovation program (grant agreement no. 101018894). I thank the ERC (101018894), the Deutsche Forschungsgemeinschaft (MA 1426/21-3) and the Simons-Moore Initiative on the Origin of Eukaryotic Cells (9743) for funding.

**Author contributions**
W.F.M. performed the calculations, made the figures and wrote the paper.

**Supplemental Table S1:** Calculated costs for synthesizing an amino acid (one letter code) in *Methanococcus* using $NH_4^+$ or $N_2$ as nitrogen source. Amino acid frequencies were calculated from the *Methanococcus thermolithotrophicus* genome sequence. As shown in Figure 1 of the main text, the estimation of amino acid frequencies in protein from the genome sequence and chemical method for *E. coli* correspond well.

**Table S1:** ATP requirements for amino acid synthesis in a $H_2$-dependent methanogen grown on $NH_4^+$ or $N_2$

| AA | mmol per 524 mg protein | Precursor, cost [ATP][c] | additional ATP | $NH_4^+$ ATP per amino acid | $NH_4^+$ mmol ATP per g cells | $N_2$ ATP per amino acid | $N_2$ mmol ATP per g cells |
|---|---|---|---|---|---|---|---|
| A | 0.255 | Pyr [0] | N [1] | 1 | 0.254 | 9 | 2.29 |
| C | 0.066 | 3PGA [2] | N [1] | 3 | 0.199 | 11 | 0.731 |
| D | 0.278 | OA [3] | N [1] | 4 | 1.115 | 12 | 3.35 |
| E | 0.397 | 2OG [4] | N [1] | 5 | 1.984 | 13 | 5.15 |
| F | 0.183 | E4P [3] PEP [2] | Chor. [1] N [1] | 7 | 1.277 | 15 | 2.74 |
| G | 0.318 | 3PGA [2] | N [1] | 3 | 0.955 | 11 | 3.50 |
| H | 0.076 | PRPP [5] | 2 PPi [4] N [3] | 12 | 0.916 | 36 | 2.75 |
| I | 0.464 | Pyr [0] Thr [7] | N [1] | 8 | 3.714 | 16 | 7.43 |
| K[a] | 0.464 | Pyr [0] Asp-SA [5] | Succ-CoA [1] N [1] | 7 | 3.249 | 23 | 10.7 |
| L | 0.431 | Pyr [0] Ac-CoA [0] | N [1] | 1 | 0.430 | 9 | 3.87 |
| M | 0.119 | Pyr [0] Asp-SA [5] | Succ-CoA [1] N [1] | 7 | 0.833 | 15 | 1.78 |
| N | 0.274 | Asp [4] | PPi [2] N [1] | 7 | 1.916 | 23 | 6.30 |
| P | 0.165 | Glu [5] | γ-Glu-P [1] | 6 | 0.988 | 14 | 2.30 |
| Q | 0.080 | Glu [5] | N via γ-Glu-P [1] | 6 | 0.482 | 22 | 1.77 |
| R | 0.164 | Glu [5] CAP [1] | N-AcGlu-P [1] N [2] PPi [2] | 11 | 1.800 | 43 | 7.04 |
| S | 0.253 | 3PGA [2] | N [1] | 3 | 0.758 | 11 | 2.78 |
| T | 0.219 | Asp [4] | Asp-P [1] HS-P [1] | 6 | 1.315 | 14 | 3.07 |
| V | 0.331 | 2 Pyr [0] | N [1] | 1 | 0.331 | 9 | 2.98 |
| W[b] | 0.032 | E4P [3] PEP [2] PRPP [5] | Chor [1] N [1] Ser [3] | 14 | 0.444 | 30 | 0.95 |
| Y | 0.191 | E4P [3] PEP [2] | Chor [1] N [1] | 7 | 1.340 | 15 | 2.87 |
|  | 4.761 |  |  |  | 24.31 |  | 74.35 |

The pathways are taken from Lengeler (39). Abbreviations: Asp-P aspartyl-4-P; HS-P homoserine-P; Chor. chorismate; Asp-SA aspartate semialdehyde; Ac-CoA. Acetyl-CoA; PPi Pyrophosphate; N-AcGlu-P *N*-acetylglutamyl-P; Chor chorismate; γ-Glu-P γ-glutamyl phosphate. [a] Methanogens use the diaminopimelate pathway (93). [b] in the final step, glyceraldehyde-3-phosphate is released, which yields 1 ATP gain, hence 14 instead of 15. [c] from Figure 2 of the main text.